\documentclass[preprint,5p]{elsarticle}

\usepackage{mathrsfs}
\usepackage{verbatim}
\usepackage{booktabs}
\usepackage[tight]{subfigure}

\usepackage[latin1]{inputenc} 
\usepackage{latexsym} 
\usepackage{graphicx} 
\usepackage{amsfonts}
\usepackage{amsmath} 
\usepackage{amssymb} 
\usepackage{parskip}
\usepackage{array}
\usepackage{color}

\usepackage{slashed}
\usepackage{epsfig,graphicx,bm,color} % Include figure files
\usepackage{dcolumn}% Align table columns on decimal point
\usepackage{bm}% bold math
\usepackage{hyperref}% add hypertext capabilities
\usepackage{breakurl}
\usepackage{soul} 

\biboptions{sort&compress}

\newcommand{\be}{\begin{equation}}
\newcommand{\ee}{\end{equation}}
\newcommand{\bea}{\begin{eqnarray}}
\newcommand{\eea}{\end{eqnarray}}

\newcommand{\lsim}{\mathrel{\mathop{\kern 0pt \rlap
  {\raise.2ex\hbox{$<$}}}
  \lower.9ex\hbox{\kern-.190em $\sim$}}}
\newcommand{\gsim}{\mathrel{\mathop{\kern 0pt \rlap
  {\raise.2ex\hbox{$>$}}}
  \lower.9ex\hbox{\kern-.190em $\sim$}}}

\newcommand{\tev}{\mbox{ TeV}}

\newcommand{\pb}{\mbox{ pb}}

\begin{document}

\begin{frontmatter}

\title{A resonance without resonance:\\
scrutinizing the diphoton excess at 750 GeV}
\tnotetext[preprints]{IFT-UAM/CSIC-15-137, DESY 15-255}

\author[kr1]{Jong Soo Kim}
\ead{jong.kim@csic.es}

\author[jr]{J\"urgen Reuter}
\ead{juergen.reuter@desy.de}

\author[kr1,kr2]{Krzysztof Rolbiecki}
\ead{rolbiecki.krzysztof@csic.es}

\author[rr]{Roberto Ruiz de Austri}
\ead{rruiz@ific.uv.es}

\address[kr1]{Instituto de F\'{\i}sica Te\'{o}rica UAM/CSIC, Madrid, 
  Spain}
  \address[jr]{DESY, Hamburg, Germany}
\address[kr2]{Institute of Theoretical Physics, University of Warsaw, Warsaw, 
  Poland}
\address[rr]{Instituto de F\'isica Corpuscular, IFIC-UV/CSIC, Valencia, Spain}

\begin{abstract}
 
Motivated by the recent diphoton excesses reported by both ATLAS and
CMS collaborations, we suggest that a new heavy spinless particle is
produced in gluon fusion at the LHC and decays   
to a couple of lighter pseudoscalars which then decay to photons. The new
resonances could arise from a new strongly interacting sector and couple
to Standard Model gauge bosons only via the corresponding
Wess-Zumino-Witten anomaly. We present a detailed recast of the newest
13 TeV data from ATLAS and CMS together with the 8~TeV data to scan
the consistency of the parameter space for those resonances.
 
\end{abstract}

\begin{keyword}
%% keywords here, in the form: keyword \sep keyword
BSM phenomenology \sep LHC
%% MSC codes here, in the form: \MSC code \sep code
%% or \MSC[2008] code \sep code (2000 is the default)

\end{keyword}

\end{frontmatter}

\section{Introduction}

After analyzing the first $13$ TeV data, the ATLAS and CMS
collaborations have reported an excess  
with respect to the background predictions in the diphoton channel
search \cite{ATLASdiphoton2015,CMSdiphoton2015}.  
ATLAS has found the most significant deviation for a mass of about
$750$ GeV, corresponding to a local  
significance of $3.64 \, \sigma$ using $3.3$ fb$^{-1}$ accumulated
data,  whereas CMS has a significance  
of $2.6 \, \sigma$ for a mass about the same as ATLAS using $2.7$
fb$^{-1}$ data. 

A simple explanation of such an excess could be through a resonance of  
a spin-0 or spin-2 particle with mass $\sim 750$ GeV that decays to
photons
\cite{Harigaya:2015ezk,Mambrini:2015wyu,Backovic:2015fnp,Angelescu:2015uiz, 
Buttazzo:2015txu,Knapen:2015dap,Nakai:2015ptz,Pilaftsis:2015ycr,Franceschini:2015kwy,DiChiara:2015vdm,Higaki:2015jag,McDermott:2015sck,Ellis:2015oso,
Low:2015qep,Bellazzini:2015nxw,Gupta:2015zzs,Petersson:2015mkr,Molinaro:2015cwg,Dutta:2015wqh,Cao:2015pto,
Matsuzaki:2015che,Kobakhidze:2015ldh,Martinez:2015kmn,Cox:2015ckc,Becirevic:2015fmu,No:2015bsn,Demidov:2015zqn,
Chao:2015ttq,Fichet:2015vvy,Curtin:2015jcv,Bian:2015kjt,Chakrabortty:2015hff,Ahmed:2015uqt,
 Agrawal:2015dbf,Csaki:2015vek,Falkowski:2015swt,Aloni:2015mxa,Bai:2015nbs}, 
while spin-1 is excluded by the Yang-Landau theorem~\cite{Landau:1948kw,Yang:1950rg}.
However, resonant production via the $s$-channel might be in tension
with bounds imposed by run-I data
\cite{Aad:2014ioa,Aad:2015mna,Khachatryan:2015qba}. This tension was
quantified in e.g.~\cite{Kim:2015ksf}.  
Though a direct production of a resonance at 750 GeV and subsequent
decay to two photons is not excluded, we propose an alternative
scenario to explain the excess via a non-resonant process. We
demonstrate our idea within the framework of strong dynamics around the TeV scale. 
The basis of the theoretical model are outlined in 
\cite{Cacciapaglia:2015nga} (some of these ideas have also been
mentioned in~\cite{Harigaya:2015ezk}) with the addition of two
composite singlet scalar 
(or pseudoscalar) particles which couple to Standard Model (SM) gauge
bosons via the Wess-Zumino-Witten anomaly~\cite{Wess:1971yu,Witten:1983tw}, and to each other via a 
trilinear coupling. Here, we assume that the lighter 750 GeV
pseudoscalar has no effective coupling  
to gluons and thus cannot be directly produced.
Therefore, the lighter pseudoscalar has to be produced in the decay of the heavy one which can be produced via gluon fusion.
In contrast to e.g.~\cite{Harigaya:2015ezk} we do not try to embed
this into a complete model, but concentrate on the minimal simplified
model that resembles a ``composite sector toy model'' including two
resonances to describe the LHC results from the 8 and 13 TeV data sets.

The parameter space of the model then consists of the masses of the
two (pseudo)scalar states, their decay constants, and their three
operator coefficients to the field strengths of the three gauge groups of
the SM. In this work we search for solutions that explain
the excess, and to determine the best fit to the data by means of a
numerical analysis.

This work is structured as follows. In Section 2 we present a 
general description of our model assumptions, whereas in Section 3 
we describe the numerical procedure used to fit the data and show
the results, and finally in Section 4 the conclusions are outlined.

%%%%%%%%%%%%%%%%%%%%%%%%%%%%%%%%%%%%%%%%%%%%%%%%
%%%%%%%%%%%%%%%%%%%%%%%%%%%%%%%%%%%%%%%%%%%%%%%%

\section{Model assumptions\label{sec:model}}

In this section, we describe in detail the assumptions for our
simplified model setup to explain the ATLAS and CMS data. We discuss
the most important phenomenological aspects of the simplified model
below. Detailed information about the underlying assumptions on
strongly interacting sectors for such a setup can be found e.g.\ in
Ref.~\cite{Cacciapaglia:2015nga,Harigaya:2015ezk}.  

In our simplified model, we consider the SM particle spectrum extended
by possible weak scale singlet spin 0 resonances. We assume that these new
resonances (and possibly also the SM-like Higgs boson) are composite
objects. However, the details of the electroweak symmetry breaking will not
affect our numerical analysis and its results and thus we do not 
discuss it any further. We assume a hidden strongly interacting
(confining) gauge group $G_N$. Two pseudoscalar resonances $\sigma$ and
$\eta$ emerge as the Nambu Goldstone bosons of the broken gauge
group $G_N$.\footnote{There are possibly more resonances, but they do not
  play a role for the moment as no further signals have been observed
  yet.} The kinetic terms of the weak singlet pseudoscalars are 
given by   
\begin{eqnarray}
\hspace*{-0.5cm} \mathcal{L}_{\rm{kin}}= \frac12 \partial_\mu\eta\partial^\mu\eta +
\frac12 \partial_\mu\sigma\partial^\mu\sigma - \frac12 m_{\eta}^2\eta^2 -
\frac12 m_{\sigma}^2\sigma^2\,.
\end{eqnarray}
Here, $m_\eta$ and $m_\sigma$ are the mass terms of the real
pseudoscalar fields $\eta$ and $\sigma$.  
We assume the following parity violating trilinear
$\sigma$-$\eta$-$\eta$ interaction term,\footnote{We assume that parity
  is explicitly violated via a nonzero $\theta$ term in the gauge
  group $G_N$. In principle, the heavier resonance $\sigma$ could also
  be scalar, without the need for CP violation in this
  interaction. However, then it needs to be a glueball in the
  confining theory which is difficult to justify why it should be so
  much heavier then the $\eta$~\cite{Cacciapaglia:2015nga}.} 
\begin{equation}
  \label{eq:trilinear}
\mathcal{L}_{\rm{trilinear}}=\lambda\,\sigma\eta\eta\,,
\end{equation}
where $\lambda$ is a real parameter of mass dimension one. In a more
general framework, all interaction terms up to mass dimension 4
consistent with our model should be included. However, since the
diphoton excess can be explained with the trilinear interaction
term only, we will omit these terms in the remainder of the letter. 
Note that such a term is the simplest assumption one can make about
such a trilinear coupling. This kind of coupling arises e.g. in the
form of a scalar-pseudoscalar-pseudoscalar coupling, $\Phi\eta\eta$,
in certain types of Little Higgs
models~\cite{Kilian:2004pp,Kilian:2006eh}. These are variants of
composite models, endowed with a certain symmetry structure. Similar
mechanisms can generate such couplings also in plain composite
models. Introducing explicit CP violation into symmetry-breaking terms
of the non-linear sector will correspondingly generate couplings of
the type Eq.~\eqref{eq:trilinear}. 

Alternatively, such couplings could arise with a different Lorentz
structure using chiral perturbation theory for composite models 
as~\cite{Gasser:1983yg,Gasser:1984gg}
\begin{eqnarray}
  && \mathcal{L} = \text{tr} \left[ \partial_\mu U \partial^\mu U^\dagger
    \left( M U + U^\dagger M\right)\right] \nonumber \\ 
  \Longrightarrow &&
  \mathcal{L}'_{\rm{trilinear}}=\lambda' \,\sigma (\partial_\mu
  \eta \partial^\mu \eta) \,.
\end{eqnarray}
Here, $U$ is the Goldstone boson non-linear field matrix, and $M$ is a
mass matrix for the underlying, condensing new fermions. Though this
leads to a different Lorentz structure, for the signal rate arguments
used in this paper, it does not change the conclusions. 

The new resonances only couple to the SM gauge bosons via the
Wess-Zumino-Witten (WZW) anomaly~\cite{Wess:1971yu,Witten:1983tw}, 
\begin{eqnarray}
\mathcal{L}_{\phi
  gg}&=&\kappa_g^\phi\frac{g_3^2}{32\pi^2}\frac{1}{F_\phi}\epsilon^{\mu\nu\rho\sigma}G_{\mu\nu}^a
G_{\rho\sigma}^a\phi,\\ 
\mathcal{L}_{\phi
  WW}&=&\kappa_W^\phi\frac{g_2^2}{32\pi^2}\frac{1}{F_\phi}\epsilon^{\mu\nu\rho\sigma}W_{\mu\nu}^i
W_{\rho\sigma}^i\phi,\\ 
\mathcal{L}_{\phi
  BB}&=&\kappa_B^\phi\frac{g_Y^2}{32\pi^2}\frac{1}{F_\phi}\epsilon^{\mu\nu\rho\sigma}B_{\mu\nu}
B_{\rho\sigma}\phi, 
\end{eqnarray}
with $\phi = \eta$ or $\sigma$. Here, $\kappa_i^\eta$,  $\kappa_i^\sigma$ and $F_\eta$,
$F_\sigma$ denote arbitrary real coefficients and pseudoscalar decay
constants, respectively. $G_{\mu\nu}$, $W_{\mu\nu}$ and $B_{\mu\nu}$
are the color, weak isospin and abelian hypercharge field strength,
and $g_3$, $g_2$ and $g_Y$ denote the corresponding 
dimensionless SM gauge couplings. The prefactors $\kappa_i^\eta$ and
$\kappa_i^\sigma$ can be explicitly calculated in a complete model,
i.e.\ if the particle content (fermions in the composite sector and
their exact quantum numbers) in the triangle loop is known~\cite{Harigaya:2015ezk,Cacciapaglia:2015nga}. However, in this work we
do not consider a particular model and assume that the coefficients
$\kappa_i^\eta$ and $\kappa_i^\sigma$ are free parameters of our
effective Lagrangian. In the following, we will assume that the
coefficients are independent and determine their values in a numerical
analysis without referring to a specific model. A possible realization
of the phenomenological model discussed here will be presented below at
the end of the section.

In this paper, we assume that the 750 GeV resonance is not directly
produced via $s$ channel, as this has already been studied in the
literature, and there is this tension with constraints from Run 1
data. In order to accomplish indirect production, we set the
corresponding anomaly coefficient to zero, $\kappa_g^\eta=0$. So its
production must occur via the heavy resonance $\sigma$ assuming that
$\sigma$ has anomaly induced couplings to the gluons.  Thus, we consider
a hierarchical scenario in order to evade the 8 TeV limits. We focus
on resonant production of the heavy singlet pseudoscalar $\sigma$ via
gluon fusion with subsequent on-shell decay into a pair of $\eta$s.  
The light pseudoscalar $\eta$ is allowed to decay into all electroweak
SM gauge bosons via the WZW mechanism (but at least photons and $Z$,
which is inevitable). Thus, we expect the following signature 
\begin{equation}
pp\rightarrow \sigma\rightarrow\eta\eta\rightarrow\gamma\gamma+X,
\end{equation}
where $X$ denotes the rest of the event. Both experiments, ATLAS and
CMS do not veto on $X$. We have listed the selection cuts from ATLAS
and CMS in Table~\ref{tab:selection13tev}. Hence, this signature can
explain the diphoton excess. The anomaly coefficients for the weak and
hypercharge group have been partially set to zero as they are
phenomenologically not relevant for the numerical analysis (in the
case of the heavy resonance), or are not allowed in order not to give
a too small branching fraction into photons (for the light resonance). 
Note that there is a certain redundancy of parameters in the
simplified model, as changes to the decay constant can within a
certain range of parameters always be emulated by changes in the
anomaly coefficient.\footnote{Note that our setup in principle also
  includes glueballs of the composite theory where, however, the
  decomposition of the operator coefficients into dimensionful
  parameters would be different due to a different power counting.} 

Our choice of anomaly coefficients
resembles the following composite model which was discussed in
Ref.~\cite{Harigaya:2015ezk}. Here, two vector-like fermions $Q_1$ and
$Q_2$ are introduced. Both vector-like fermions are in the fundamental
representation of the strong gauge group $G_N$  
with a dynamical scale $\Lambda$. Here, we identify $G_N$ with an asymptotically free SU(N) gauge theory. The charge assignments of $Q_1$ and $Q_2$ under
$SU(N)\times$SU(3)$_C\times$SU(2)$_L\times$U(1)$_Y$ are $(N,1,2,a)$
and $(N,\bar 3,1,b)$, respectively. $N$ denotes the fundamental
representation of the $SU(N)$. Now, we assume that $m_1<\Lambda<m_2$. 
At low energies, we have the following condensate
\begin{equation}
\eta\sim\langle Q_1\bar Q_1 \rangle,
\end{equation}
where $\eta$ is a triplet under SU(2)$_L$.  
The only allowed ano\-maly induced coupling of $\eta$ is given by the
following term 
\begin{equation}
\mathcal{L}\propto \eta^b W^{b\mu\nu} \tilde B_{\mu\nu}.
\end{equation}
Since the (pseudo)scalar $\eta^b$ cannot couple to gluons, it cannot be
produced directly. However, heavy resonances with $Q_1$ pairs can be
produced which then decay into the lighter pseudoscalar via
Eq.~\eqref{eq:trilinear}. The latter can then decay to two
photons. Using more than two underlying fundamental fermions of the
confining gauge group $SU(N)$ with appropriate quantum numbers, more
general scenarios can be constructed.

%Once a specific composite model is investigated,
%this has to be carefully disentangled from each other.
%%%%%%%%%%%%%%%%%%%%%%%%%%%%%%%%%
%%%%%%%%%%%%%%%%%%%%%%%%%%%%%%%%%

\section{Numerical results}

In the following, we first briefly
discuss the numerical tools and then describe our numerical framework.
Finally, we will discuss our results.

%%%%%%%%%%%%%%%%%%%%%%%%%%%%%%
%%%%%%%%%%%%%%%%%%%%%%%%%%%%%%%

\subsection{Numerical tools}

\begin{table}
\centering \renewcommand{\arraystretch}{1.3}
\scalebox{0.8}{
\begin{tabular}{|c|c|}\hline
 ATLAS & CMS \\
\hline
\hline
$p_T(\gamma)\ge$25 GeV  &
$p_T(\gamma)\ge$75 GeV \\ 
\hline
$|\eta^{\gamma}|\le2.37$ & $|\eta^{\gamma}|\le 1.44$
 or $1.57 \le |\eta^{\gamma}| \le 2.5$ 
\\ 
& at least one $\gamma$ with $|\eta^{\gamma}|\le1.44$\\
\hline
$E_T^{\gamma_1}/m_{\gamma\gamma}\ge0.4$,
$E_T^{\gamma_2}/m_{\gamma\gamma}\ge0.3$ & $m_{\gamma\gamma}\ge230$ GeV \\ \hline
\end{tabular}}
\caption{Selection cuts of the 13 TeV ATLAS/CMS diphoton searches
  \cite{ATLASdiphoton2015,CMSdiphoton2015}. %%\RR{Checked}
\label{tab:selection13tev} }
\end{table}

We have implemented the model discussed in Section~\ref{sec:model} with the program {\tt FeynRules 2.3.13}
\cite{Alloul:2013bka} and created a {\tt UFO} output  
\cite{Degrande:2011ua} for the numerical studies. 
Parton level events were generated with {\tt Madgraph 2.3.3}
\cite{Alwall:2014hca} interfaced with {\tt Pythia 6.4}
\cite{Sjostrand:2006za} for the parton shower and hadronization.
Branching ratios and cross sections have been cross-checked with an
independent numerical implementation of the simplified model 
into {\tt WHIZARD 2.2.8}~\cite{Kilian:2007gr,Moretti:2001zz,Christensen:2010wz}.
We have implemented the 8 and 13 TeV diphoton searches from ATLAS and
CMS \cite{ATLASdiphoton2015,CMSdiphoton2015}  
into the {\tt CheckMATE 1.2.2} framework \cite{Drees:2013wra} with its
{\tt AnalysisManager} \cite{Kim:2015wza}.  \linebreak[4]
{\tt CheckMATE 1.2.2} is based on the fast detector simulation {\tt
  Delphes 3.10} \cite{deFavereau:2013fsa} with heavily  
modified detector tunes %of the ATLAS detector  
and it determines the number of expected signal events passing the
selection cuts of the particular analysis. The selection cuts  
for both ATLAS and CMS analyses are shown in Table~\ref{tab:selection13tev}.

%%%%%%%%%%%%%%%%%%%%%%%%%%%%%%%%
%%%%%%%%%%%%%%%%%%%%%%%%%%%%%%

\subsection{Scan procedure}
In order to find values of parameters that provide a good description
of data we performed a scan in $\lambda$, $\kappa_g^\sigma$ and the mass of heavier resonance, $m_{\sigma}$, in the 
ranges displayed in Table~\ref{tab:parameters}. We simulated pair
production of $\eta$ states via resonant $s$-channel $\sigma$
exchange.\footnote{This neglects possible box diagram contributions to
  $\eta$ pair production which, however, are equal to zero if the QCD
  WZW anomaly of the $\eta$ current vanishes.} All
decay modes of $\eta$ were included in the simulation.  For each point
the number of events passing experimental selections in our simulation
is compared to the number of events reported by the LHC
collaborations, see Table~\ref{tab:result}. The expected number of background events is extracted
from the respective publication. Because the experiments did not
clearly define signal regions, we performed a fit in the invariant mass
window $ 700 < m_{\gamma\gamma} < 800$~GeV. For the CMS search we
split the events into the barrel (EBEB) and end-cap (EBEE) regions, following the
collaboration's procedure. Finally, the 8~TeV searches are used solely
as a consistency check in order to see if the parameter points were
not excluded during the previous run.

\begin{table}[t!]
\begin{center}
\scalebox{0.89}{
\begin{tabular}{|c|c|c|}
\hline
Parameter & Description & Value or range \\  %& Best Fit Point\\
\hline
$m_\sigma$& mass of heavier resonance & $[1.5\ {\rm TeV},\, 2.5\ {\rm TeV}]$ \\
$m_\eta$ & mass of lighter resonance & $750\ {\rm GeV}$ \\
%$\lambda$ & dimensionfull $\eta\sigma\sigma$ & $[0.5 \ {\rm TeV},\, 2.5 \ {\rm TeV}]$ &\\
$\lambda$ & dimensionfull $\eta\sigma\sigma$ & $[0. \ {\rm TeV},\, 1.5 \ {\rm TeV}]$ \\
$F_\eta$ & $\eta$ decay constant & $1 \ {\rm TeV}$ \\
$F_\sigma$ & $\sigma$ decay constant & $1\ {\rm TeV}$ \\
$\kappa_g^\eta$ & anomaly coefficient  & $0$ \\
$\kappa_W^\eta$ & anomaly coefficient & $0$ \\
$\kappa_B^\eta$ & anomaly coefficient & $1$ \\
%$\kappa_g^\sigma$ & anomaly coefficient & $[5, \, 25]$ &\\
$\kappa_g^\sigma$ & anomaly coefficient & $[0, \, 15]$ \\
$\kappa_W^\sigma$ & anomaly coefficient & $0$ \\
$\kappa_B^\sigma$ & anomaly coefficient & $0$ \\
\hline
\end{tabular}}
\end{center}
\caption{Variable input parameters of our pseudoscalar scenario and 
  the range over which these parameters are scanned to find the best fit
  point solution.}  
 \label{tab:parameters} 
\end{table}

\begin{figure}[t!]
\begin{center}                                                                                                                                  
\includegraphics[width=0.45\textwidth]{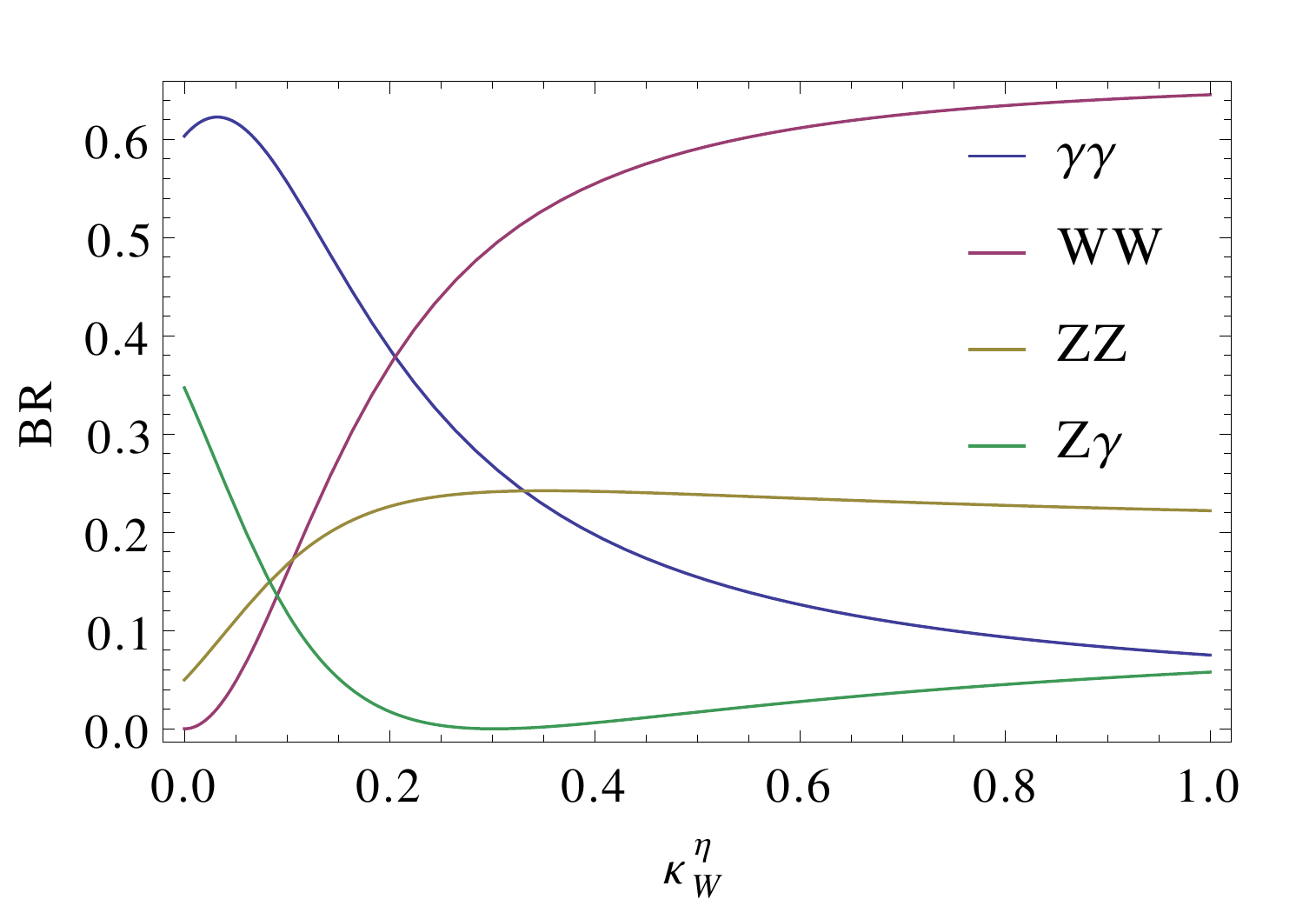} 
\caption{The branching ratios for the decays of $\eta$ as a function of $\kappa_W^\eta$ with $\kappa_B^\eta = 1$.
\label{fig:brs}}
\end{center}
\end{figure}

Clearly, the pseudoscalar sector is parametrized by several {\em a priori} free
parameters, as can be seen in Table~\ref{tab:parameters}. For
simplicity, in the current analysis we set some of them to zero, therefore our
heavy pseudoscalar couples only to gluons through the anomaly while the
light one only to $B$. The light pseudoscalar will still have other decay
modes to $ZZ$ and $Z\gamma$. Once the $\kappa_W^\eta$ coupling is
allowed, additional decay modes to $WW$ pairs will be open. We show
the decay pattern of $\eta$ in Figure~\ref{fig:brs}. This will provide
a distinctive feature at colliders: the diphoton pairs will be
often accompanied by jets, leptons or missing transverse
energy. Depending on the actual mass hierarchy between scalars, the
$\gamma\gamma$ pair can have a significant transverse momentum, which
will eventually vanish close to the threshold for the decay $\sigma
\to \eta \eta$. This feature together with an additional activity in
the event can serve as a way to probe this type of models
soon~\footnote{We note that according to the CMS speaker's statement
  during the seminar presenting those results at CERN, no difference
  between the peak and side-band regions was observed. Given the low
  number of events and lack of details in the conference 
  note~\cite{CMSdiphoton2015} we do not take this remark as
  conclusive. }.

The fit was performed with the $\chi^2$ test statistics. Namely, 
\begin{eqnarray}
 \chi_i^2 = \frac{(n_i-\mu_i)^2}{\sigma^2_{i,{\rm stat}}+\sigma_{i,b}^2}\;, 
\end{eqnarray}
where
\begin{eqnarray}
 \mu_i = \mu_{i,b}+\mu_{i,s}\;.
 \end{eqnarray}
Here, $n_i$ is the number of observed events, $\mu_{i,b}$ is the
expected number of background  
events, $\mu_{i,s}$ is the expected number of signal 
events, $\sigma_{i,{\rm stat}}$ and $\sigma_{i,b}$ are the statistical
and systematic uncertainty  
on the expected number of background events for each signal region,
where the index $i$ runs over the $i = $ ATLAS, CMS EBEB, CMS EBEE
selections. The systematic errors combine the experimental ones as
given by the collaborations and additionally $10\%$ error on the
CheckMATE event yield. In any case, the total error is dominated by
the statistical errors due to the low number of events. We assume
that all errors are uncorrelated. The signal regions are defined as
$700 < m_{\gamma\gamma} < 800$~GeV.

%%%%%%%%%%%%%%%%%%%%%%%%%%%%%
%%%%%%%%%%%%%%%%%%%%%%%%%%%%%

\subsection{Discussion}

As explained above, we have performed a scan as a function of the couplings $\lambda$, $\kappa^\sigma_g$ and the mass $ m_{\sigma}$ while keeping the other parameters fixed as it is  shown in Table~\ref{tab:parameters}. This has been done with the ATLAS and CMS searches and in the following we combine the data of both  experiments. 

\begin{figure*}[t!]
\begin{center}                                                                                                                                    
\includegraphics[width=0.32\textwidth]{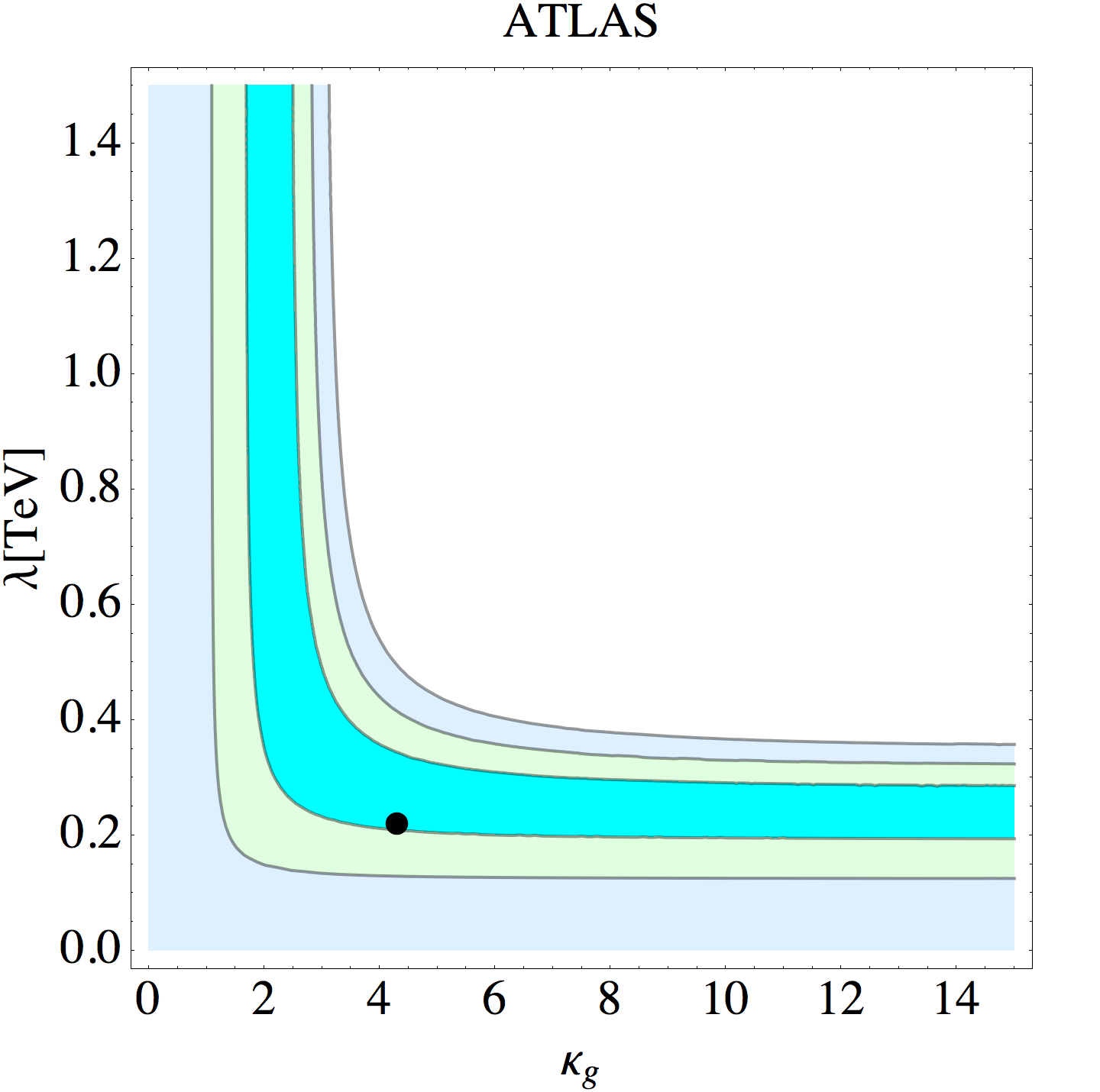} 
\includegraphics[width=0.32\textwidth]{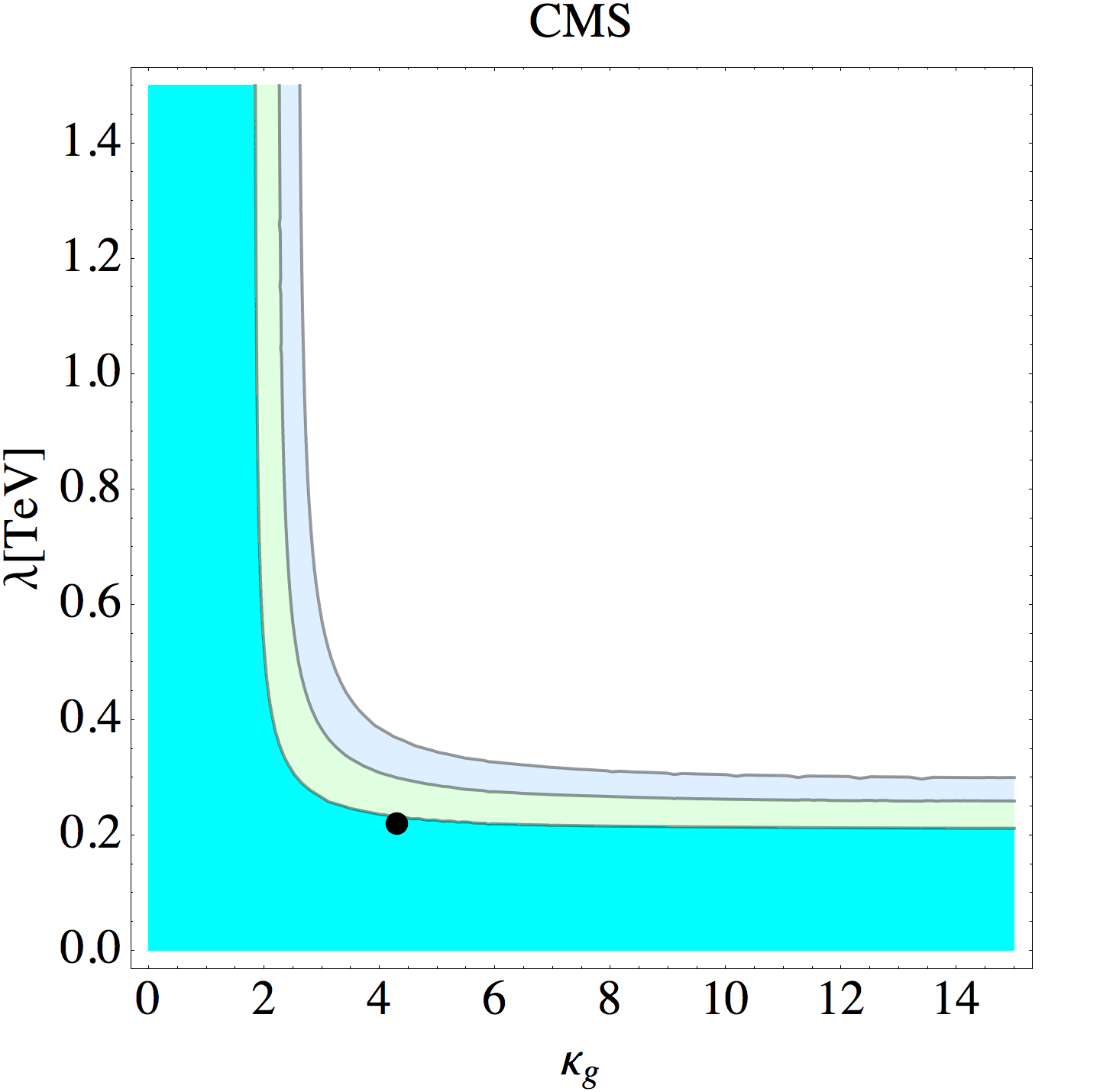} 
\includegraphics[width=0.32\textwidth]{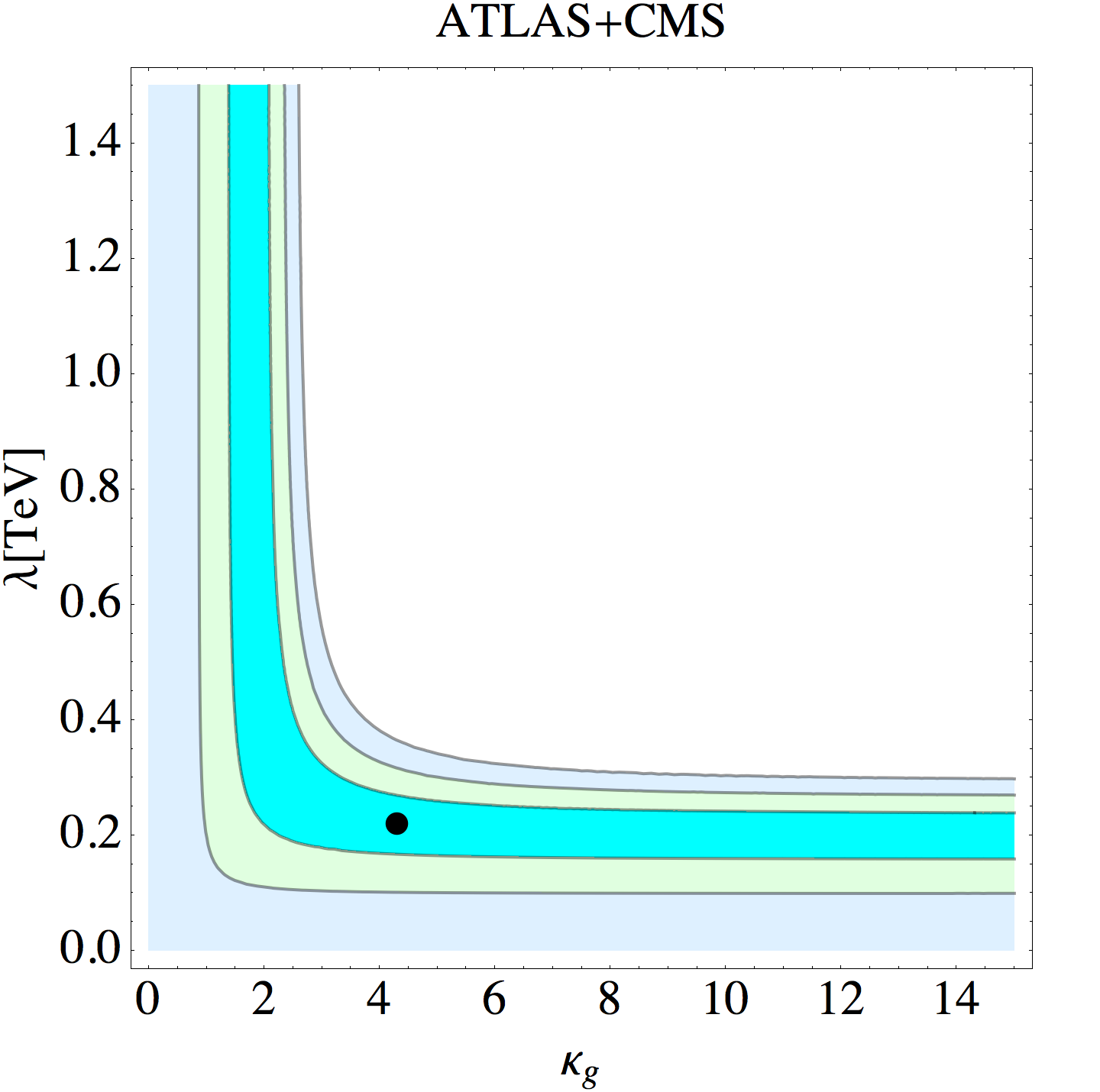} 
%\vspace{-0.75cm}
%a) \hspace{0.\textwidth} b)\hspace{0.45\textwidth}
\caption{The distribution of the $\chi^2$ test as a function of the dimensionful coupling $\kappa^\sigma_g$ and $\lambda$  for ATLAS \cite{ATLASdiphoton2015}, CMS \cite{CMSdiphoton2015} and both experiments combined with $m_{\sigma}$ set to $1.75 \tev$. The colors and contours denote: cyan $\Delta\chi^2 = 1$ above minimum; light green $\Delta\chi^2 = 4$  above minimum; and light blue $\Delta\chi^2 = 9$  above minimum. The dots represent sample best-fit point: $\lambda = 0.22 \tev$, $\kappa^{\sigma}_g =  4.3$, $m_{\sigma} = 1.75 \tev$. }
\label{fig:chi2_kg_lambda}
\end{center}
\end{figure*}

Figure~\ref{fig:chi2_kg_lambda} shows the $\chi^2$ as a function of
$\kappa^\sigma_g$ and $\lambda$ (left panel for ATLAS data alone,
middle panel for CMS data alone and right panel for their combination). The contours are plotted for $\Delta \chi^2 = 1,\,4,\,9$  above minimum respectively.
It is interesting to notice that the $\chi^2$ follows a hyperbolic shape in the $\lambda$, $\kappa^{\sigma}_g$ plane since as it is expected there is a degeneracy among the two couplings. Namely an increase in the $\kappa^\sigma_g$ coupling enhances the production rate of the $\sigma$ field via gluon-fusion which is
compensated by a decrease of the $\lambda$ coupling which affects the decay branching fraction of $\sigma$ decaying to a pair of $\eta$'s. The black dots represent one of the low $\chi^2$ points: $\lambda = 0.22 \tev$, $\kappa^{\sigma}_g =  4.3$, $m_{\sigma} = 1.75 \tev$.

\begin{figure*}[t!]
\begin{center}                                                                                                                                   
\includegraphics[width=0.32\textwidth]{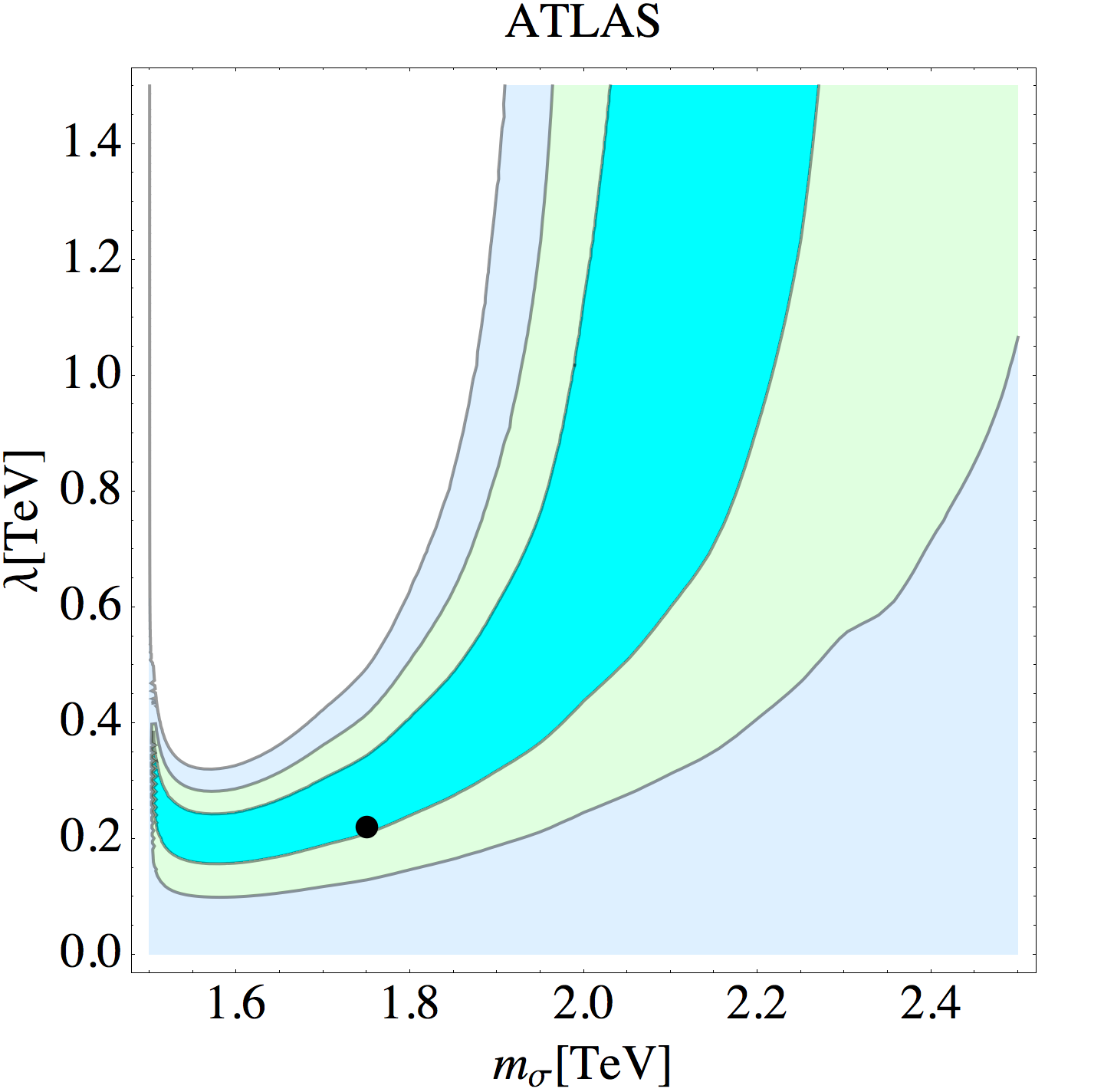} 
\includegraphics[width=0.32\textwidth]{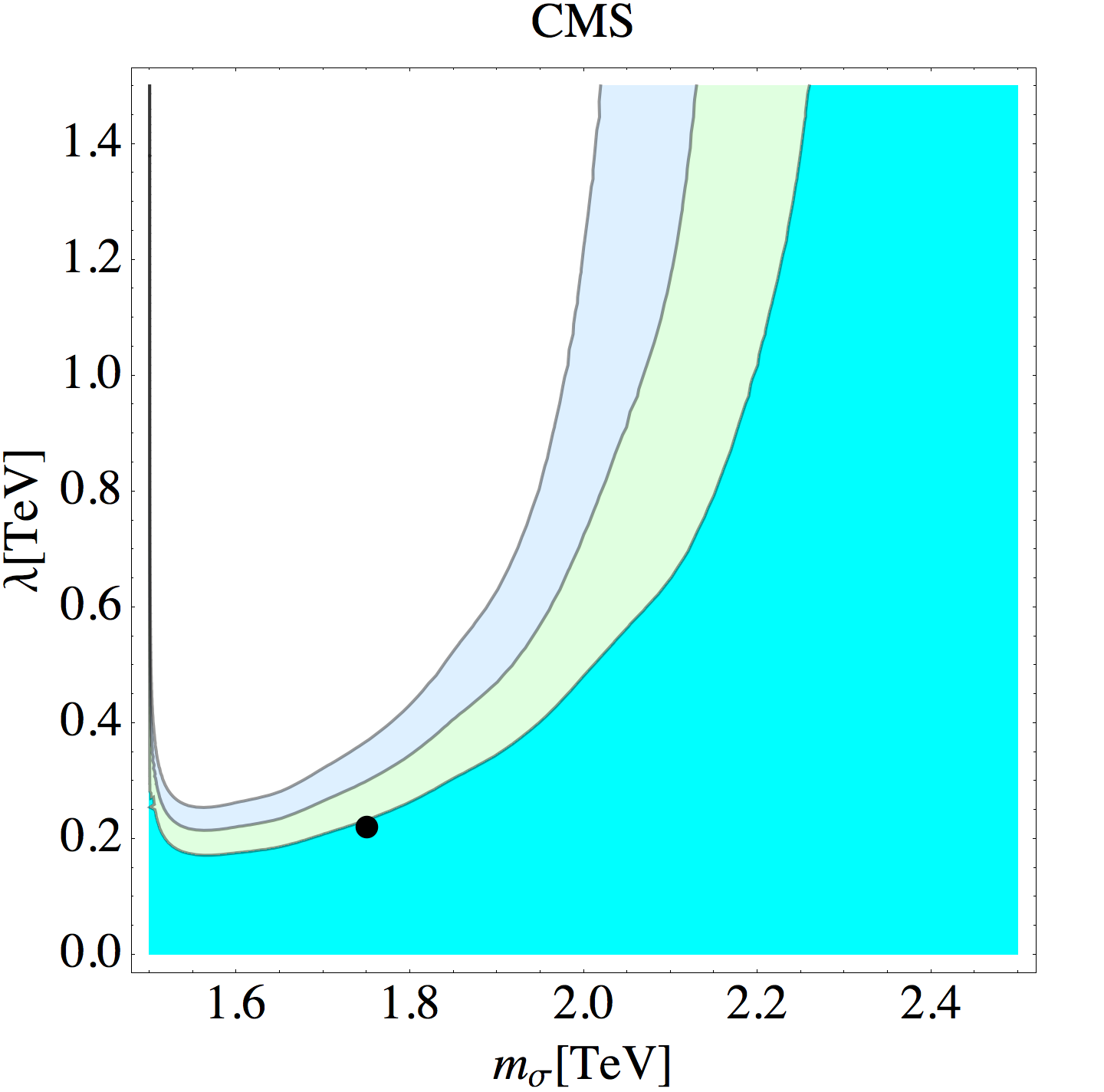} 
\includegraphics[width=0.32\textwidth]{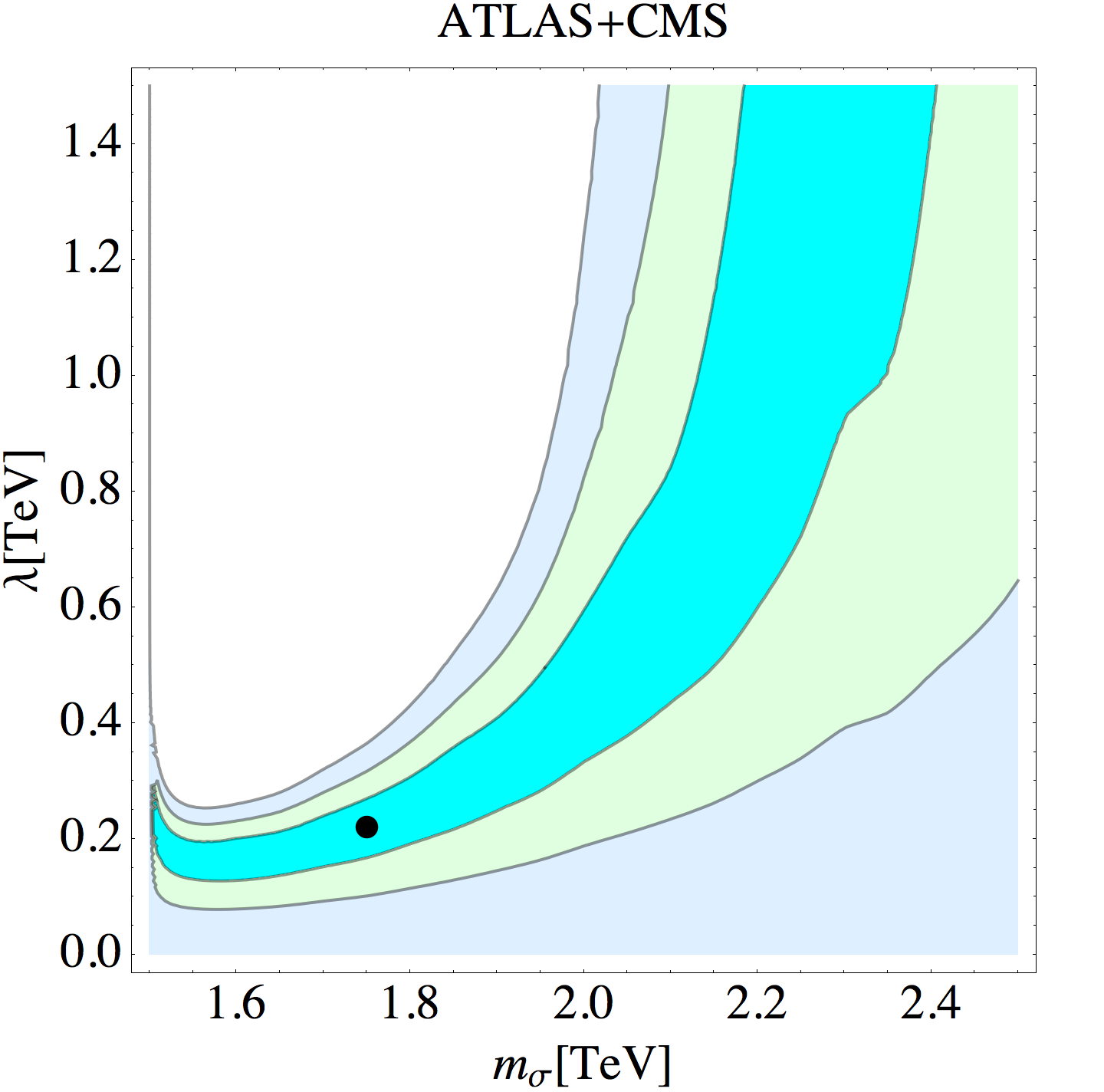} 
%\vspace{-0.75cm}
%a) \hspace{0.\textwidth} b)\hspace{0.45\textwidth}
\caption{The distribution of the $\chi^2$ test as a function of the
  dimensionful coupling $m_{\sigma}$ and $\lambda$ for ATLAS
  \cite{ATLASdiphoton2015}, CMS \cite{CMSdiphoton2015} and both
  experiments combined with $\kappa^{\sigma}_g$ set to $4.3$. The
  colors and contours denote: cyan $\Delta\chi^2 = 1$ above minimum;
  light green $\Delta\chi^2 = 4$  above minimum; and light blue
  $\Delta\chi^2 = 9$  above minimum. The dots represent the sample
  best-fit point: $\lambda = 0.22 \tev$, $\kappa^{\sigma}_g =  4.3$,
  $m_{\sigma} = 1.75 \tev$. } 
\label{fig:chi2_msigma_lambda}
\end{center}
\end{figure*}

\begin{figure*}[t!]
\begin{center}
\includegraphics[width=0.32\textwidth]{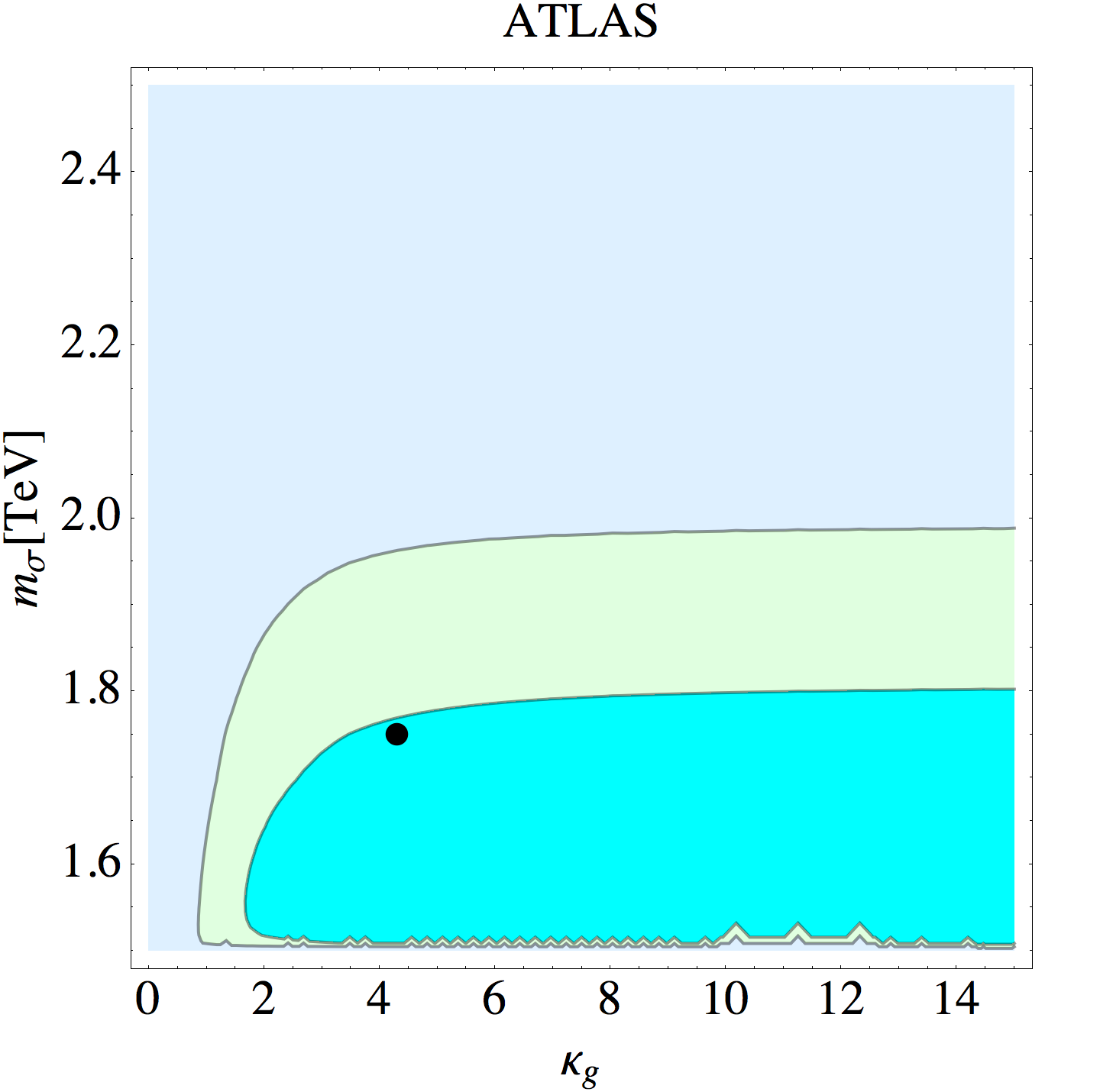} 
\includegraphics[width=0.32\textwidth]{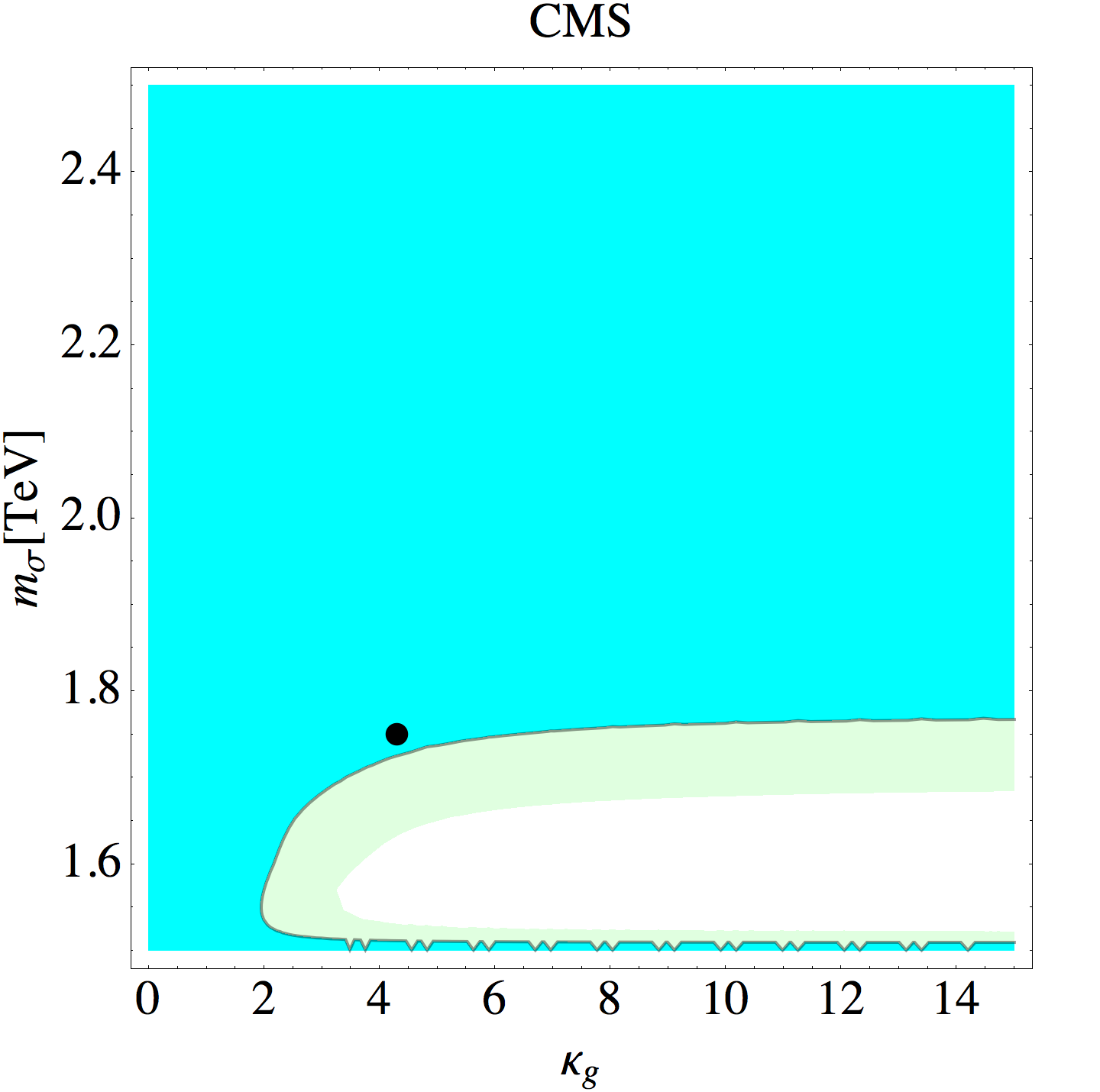} 
\includegraphics[width=0.32\textwidth]{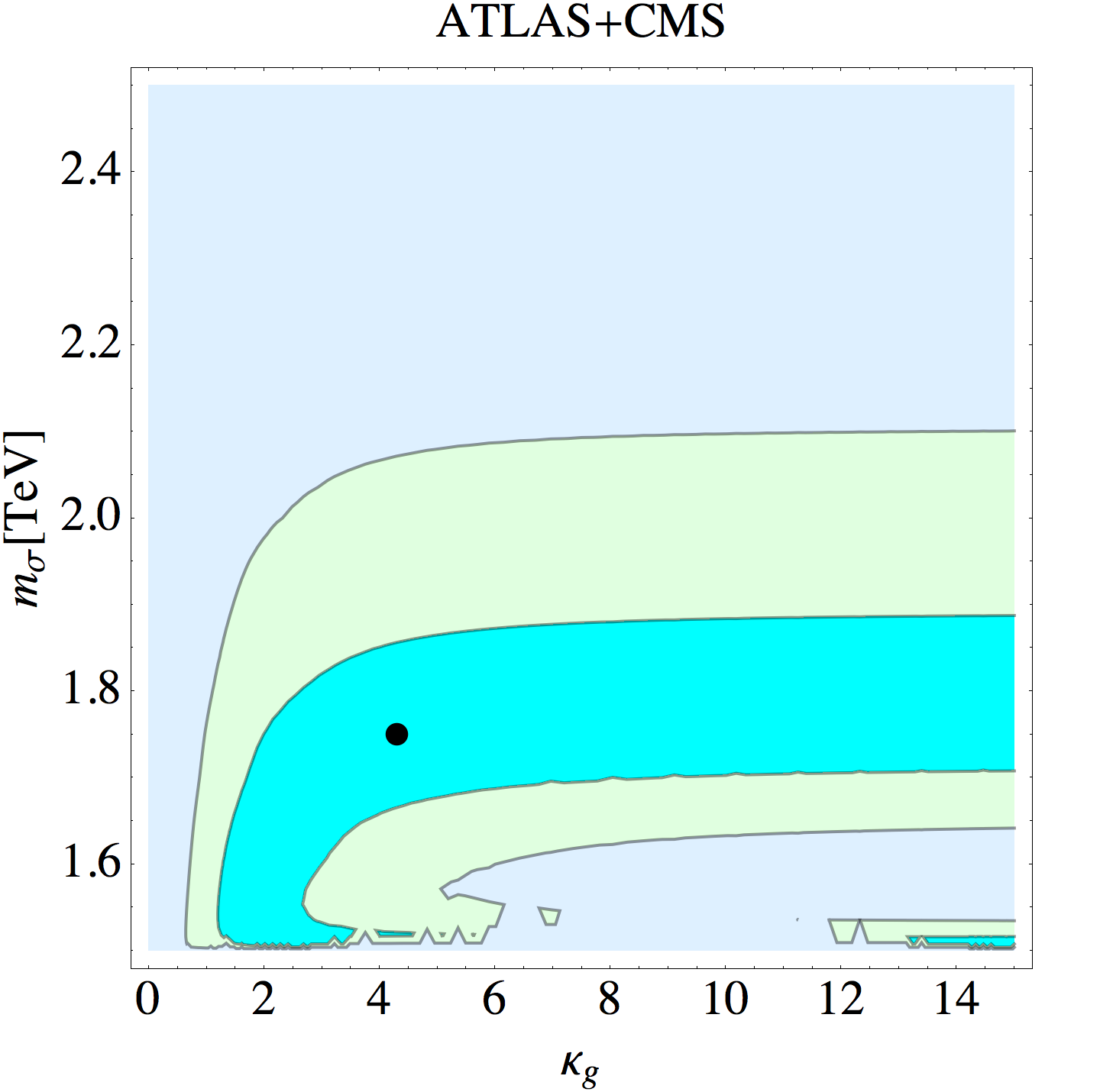} 
%\vspace{-0.75cm}
%a) \hspace{0.\textwidth} b)\hspace{0.45\textwidth}
\caption{The distribution of the $\chi^2$ test as a function of the
  dimensionful coupling $\kappa^\sigma_g$ and $m_{\sigma}$ for ATLAS
  \cite{ATLASdiphoton2015}, CMS \cite{CMSdiphoton2015} and both
  experiments combined with $\lambda$ set to $0.22 \tev$. The colors
  and contours denote: cyan $\Delta\chi^2 = 1$ above minimum; light
  green $\Delta\chi^2 = 4$  above minimum; and light blue
  $\Delta\chi^2 = 9$  above minimum. The dots represent the sample
  best-fit point: $\lambda = 0.22 \tev$, $\kappa^{\sigma}_g =  4.3$,
  $m_{\sigma} = 1.75 \tev$. } 
\label{fig:chi2_kg_meta}
\end{center}
\end{figure*}

Figure~\ref{fig:chi2_msigma_lambda}  shows the $\chi^2$ as a function
of $m_{\sigma}$ and $\lambda$ (again, left panel for ATLAS data alone,
middle panel for CMS data alone and right panel for their
combination). The contours are plotted for $\Delta \chi^2 = 1,\,4,\,9$  above minimum respectively. Finally,
Figure~\ref{fig:chi2_kg_meta}  shows the $\chi^2$ as a function of
$\kappa^\sigma_g$ and $m_{\sigma}$, with the black dot defined as
above. The preferred values of parameters lie close to the border of
1-$\sigma$ for each of the experiments. Clearly CMS is more
consistent with the no signal hypothesis, while ATLAS prefers higher
event yields.

Because we effectively only have one observable it is not surprising that there is a continuum of points with minimum $\chi^2$ value. The combined analysis gives the $\chi^2$ value at the minimum of $1.8$.
%best fit point:
%
%\begin{eqnarray}\label{eq:bestfit}
%\lambda &=&  0.41\ \mathrm{TeV}\,, \nonumber \\
%\kappa^{\sigma}_g &=&  4.9\,,  \\
%\chi^2 &=& 2.15\,. \nonumber
%\end{eqnarray}
%
This should be compared to the SM-only hypothesis which yields  $\chi^2_{\mathrm{SM}}=8.8$. Finally, we provide the expected and observed numbers of events for each signal region in Table~\ref{tab:result}.

\begin{table}[t!]
\begin{center}
\scalebox{0.89}{
\begin{tabular}{|c|c|c|c|c|}
\hline
signal region & observed & background & best fit & $\Delta \chi^2$  \\  %& Best Fit Point\\
\hline
ATLAS & $28$ & $11.4\pm 3$ & $12.0 $ & $0.56$ \\
\hline
CMS EBEB & $14$  & $9.5 \pm 1.9$  & $8.1$  & $0.74$\\
\hline
CMS EBEE & $16$  & $18.5 \pm 3.7$  & $1.3$  & $0.48$\\
\hline
\end{tabular}}
\end{center}
\caption{The number of events for each of the signal regions: observed, SM background, our 'best fit' according to the simulation results and the $\Delta \chi^2$ contribution. 'EBEB' denotes the signal region with both photons in the barrel while 'EBEE' the signal region with one photon in the end-cap. 'Best fit' point input: $\lambda = 0.22 \tev$, $\kappa^{\sigma}_g =  4.3$, $m_{\sigma} = 1.75 \tev$.  \label{tab:result} }
\end{table}

As already mentioned, because we effectively have one observable
the degeneracy in the preferred parameters cannot be resolved at
this point. The degeneracy could be broken by an observation of a
peak in the dijet mass spectra from the decay $\sigma \to gg$. Since
in the range of masses $m_{\sigma}$ studied here the results are
consistent with the background-only
hypothesis~\cite{Khachatryan:2015dcf,ATLAS:2015nsi}, one should also
take into account an additional constraint from these
searches. CMS~\cite{Khachatryan:2015dcf} provides explicit limits for
the digluon final state. For our sample point at $m_{\sigma} = 1.75
\tev$, the upper limit on the visible dijet cross section is $1.8
\pb$. Taking into account the efficiency of $~60\%$ the truth cross
section should be less than $\sim 3 \pb$. This implies
$\kappa^\sigma_g \lesssim 25$. 

Finally, we comment on the visibility of the diphoton signal at
8~TeV~\cite{Aad:2014ioa,Aad:2015mna,Khachatryan:2015qba,CMS-PAS-EXO-12-045}. 
The cross section, assuming the gluon-gluon production 
mode~\cite{Higgs:Cross}, would be a factor 15 smaller. On the other
hand, the luminosity recorded at 8~TeV was some 6 times larger than
at 13~TeV in case of ATLAS. For the sample point from
Table~\ref{tab:result} the expected event yield would be $4.8$
events in the ATLAS search~\cite{Aad:2015mna}, while the observed
(expected) $95\%$ CL upper limit is $23.1$ $(21.9)$. Similarly, in
the CMS case we find the 8~TeV constraints easily fulfilled.

%%%%%%%%%%%%%%%%%%%%%%%%%%%%%%%%%%
%%%%%%%%%%%%%%%%%%%%%%%%%%%%%%%%%

\section{Conclusions}

In this work we present a model based on composite states that fits
well the diphoton excess observed by the ATLAS and CMS collaborations
which points to the existence of a resonance of mass of about $750$ GeV. 

The mechanism consists in the production of a heavy pseudoscalar via
gluon fusion with a mass in a range $1.5$--$2.5$~TeV, which  decays to a pair of
lighter pseudoscalars with a mass of about $750$~GeV that
finally decay to photons. While both pseudoscalars couple to the SM gauge
bosons via the WZW anomaly, the heavy pseudoscalar couples to the light
ones via a dimensionful trilinear coupling which is allowed by the
theory.   

%Our best fit point from the combined analysis of the ATLAS and CMS
%data is given by  $\lambda =  0.41\ \mathrm{TeV}$ and
%$\kappa^{\sigma}_g =  4.9$. the $\chi^2$ 
We find a specific direction in the parameter space of the
  simplified model that minimizes $\chi^2$ 
from the combined analysis of the ATLAS and CMS data at the value of
$1.8$ and the models markedly improves the SM-only value of $8.8$. It
is also consistent with the 8~TeV searches. Its distinctive feature
in the collider experiments compared to the direct $s$-channel
resonance would be a non-trivial $p_T$ spectrum of the diphoton pairs
and the presence of additional jets, leptons or missing energy,
depending on the decays of gauge bosons produced in the opposite decay
chain, which makes it easily testable with more data from the upcoming
run. 

%%%%%%%%%%%%%%%%%%%%%%%%%%%%%%%%%%%%%%%%
%%%%%%%%%%%%%%%%%%%%%%%%%%%%%%%%%%%%%%%%

\section*{Acknowledgments}
We would like to thank Sascha Caron for useful discussions. RRdA is supported by the Ram\'on y Cajal program of the Spanish
MICINN and also thanks the support of the Spanish MICINN's
Consolider-Ingenio 2010 Programme  under the grant MULTIDARK
CSD2209-00064, the Invisibles European ITN project
(FP7-PEOPLE-2011-ITN, PITN-GA-2011-289442-INVISIBLES and the  
``SOM Sabor y origen de la Materia" (FPA2011-29678) and the
``Fenomenologia y Cosmologia de la Fisica mas alla del Modelo Estandar
e lmplicaciones Experimentales en la era del LHC" (FPA2010-17747) MEC
projects.  KR and JSK has been partially supported by the MINECO (Spain) under contract FPA2013-44773-P; 
Consolider-Ingenio CPAN CSD2007-00042; the Spanish MINECO Centro de
excelencia Severo Ochoa  Program under grant SEV-2012-0249; and by
JAE-Doc program. 
J.R.R wants to thank for the hospitality at the IFT of the Universidad
Autonoma de Madrid, where this work has been initiated.

%%%%%%%%%%%%%%%%%%%%%%%%%%%%%%%%
%%%%%%%%%%%%%%%%%%%%%%%%%%%%%%%%

\bibliography{diphoton_rev1}
\bibliographystyle{utphys}

\end{document}